# Ask and You Shall be Served: Representing & Solving Multi-agent Optimization Problems with Service Requesters and Providers

Maya Lavie, Tehila Caspi, Omer Lev, and Roie Zivan
Ben-Gurion University of the Negev
Be'er Sheva, Israel
{laviem,caspite}@post.bgu.ac.il,{omerlev,zinvar}@bgu.ac.il

## ABSTRACT

In scenarios with numerous emergencies that arise and require the assistance of various rescue units (e.g., medical, fire, & police forces), the rescue units would ideally be allocated quickly and distributedly while aiming to minimize casualties. This is one of many examples of distributed settings with service providers (the rescue units) and service requesters (the emergencies) which we term *service oriented settings*. Allocating the service providers in a distributed manner while aiming for a global optimum is hard to model, let alone achieve, using the existing Distributed Constraint Optimization Problem (DCOP) framework. Hence, the need for a novel approach and corresponding algorithms.

We present the Service Oriented Multi-Agent Optimization Problem (SOMAOP), a new framework that overcomes the shortcomings of DCOP in service oriented settings. We evaluate the framework using various algorithms based on auctions and matching algorithms (e.g., Gale Shapely). We empirically show that algorithms based on repeated auctions converge to a high quality solution very fast, while repeated matching problems converge slower, but produce higher quality solutions. We demonstrate the advantages of our approach over standard incomplete DCOP algorithms and a greedy centralized algorithm.

## 1 INTRODUCTION

Advances in computation and communication have resulted in realistic distributed applications in which people interact with technology to reach and optimize mutual goals, such as saving lives in disaster response [28] and maximizing user satisfaction while minimizing energy usage in smart homes [14, 30]. Thus, there is a growing need for optimization methods to support decentralized decision making in complex multi-agent systems. Many of these systems share the underlying structure of a *service oriented system*, which includes two sets of agents: one set of agents that can provide services and the other of agents that require services to be provided.

Consider, for example, a disaster rescue scenario, where rescue units (medical personnel, fire fighters, police, etc.) need to coordinate their actions to save as many people as possible from numerous disaster sites. This coordination problem is particularly challenging due to the following characteristics: **1) Optimization of a Global Objective:** the various rescue units need to work together as a team towards a common goal (e.g., saving as many victims as possible). **2) Decentralized Coordination:** often there is no centralized entity that coordinates agents, but rather a diverse set of agents (e.g., medical personnel, fire fighters and disaster site coordinators) making personal coordination decisions. While we use the disaster rescue scenario as a motivating setting throughout this paper, these factors are present in a much larger class of multi-agent coordination problems.

A common approach to solve these types of problems is to model them as *distributed constraint optimization problems* (DCOPs), where decision makers are modeled as cooperative *agents* that assign *values* to their *variables* [13, 22, 29, 32]. The goal in a DCOP is to optimize a global objective in a decentralized manner. The global objective is decomposed into constraints that define the utility agents derive (or costs they incur) from combinations of assignments to variables [5, 8, 17]. This model captures how a rescue unit (an agent) with a schedule (a variable) is assigned a disaster site to go to (a value for the variable), with the goal of saving as many victims as possible (the global objective). For each combination of assignments of disaster sites to the police units' schedules, a (possibly) different number of victims will be saved (the utility).

In DCOP algorithms, agents exchange messages, communicating selected value assignments or their estimated utilities. The information received by an agent is used to adjust its variable assignments. The local quality of the assignments they select is measured according to the constraints they are subject to.

If we examine the properties of the service oriented systems described above, it is apparent that the DCOP model does not naturally apply to them. On the contrary, in many of them, the constraints are defined by entities (e.g., disaster site coordinators) different from the agents making the decisions (e.g., rescue units). These entities require a service to be performed, but they do not assign variables. Rather, they are affected by the consequences of the decisions made by the agents performing the actions. Thus, while the solution is determined by the set of the "original" DCOP agents (the ones assigning variables, e.g., rescue units), the quality of the solution (i.e., the global utility derived from it) is measured according to the satisfaction of the service requiring agents (e.g., disaster site coordinators) from the services provided to them.

In our disaster response example, consider the ambulances that are required to evacuate casualties from disaster sites to hospitals. The number of casualties and the severity of their wounds in each disaster site determines the utility derived from evacuating them to hospitals (e.g., not much utility in evacuating people with very minor wounds). To use the standard DCOP model for solving this problem, we would have the ambulances hold complete and coherent information regarding *all disaster sites* they can drive to and exchange messages with *all rescue units* (e.g., ambulances, police units and fire fighters) that can attend to casualties from the same sites (neighboring units). Moreover, to calculate the utility that they would derive from each decision they make, the ambulances would require knowledge of *all assignments made by neighboring units* and the utility (or cost) of *all constraints* representing the outcome of each possible combination of their assignments. Such a modelling requires agents to have detailed knowledge on almost all other agents, defeating the purpose of a distributed setting.

The fact that the dominant model used to represent and solve multi-agent optimization problems seems deficient for so many distributed realistic applications is what motivates this work. We propose an alternative abstract model, *Service Oriented Multi-Agent Optimization Problem* (SOMAOP). In contrast to standard DCOP, SOMAOP offers a paradigm for multi-agent optimization that can handle service oriented settings. In this model, agents are divided into two sets: service requesters (SRs) and service providers (SPs). This approach allows us to adopt (and adapt) existing AI and OR centralized methods for assigning service providers to service requesters. Thus, in this paper we:

(1) Present the SOMAOP model.
(2) Propose algorithms based on auctions and matching algorithms for solving SOMAOPs.
(3) Conduct empirical comparison between various SOMAOP algorithms, and empirically show that algorithms based on repeated auctions converge to a high quality solution very fast, while repeated matching problems converge slower, but produce higher quality solutions.
(4) Compare SOMAOP algorithms to DCOP algorithms in solving service oriented problems.

Our results demonstrate that the SOMAOP model allows the use of algorithms that converge fast to high quality solutions while maintaining the problem's distributed structure and without requiring complete and coherent information to be held by the service providing agents.

## 2 SERVICE ORIENTED MULTI-AGENT OPTIMIZATION

The Service Oriented Multi-Agent Optimization Problem (SOMAOP) is a multi-agent problem in which there is a clear distinction between two disjoint sets of agents: *service requesting* agents (SRs) and *service providing* agents (SPs). We create a bipartite graph with the SRs and SPs as the nodes. Each service providing agent (SP) is connected by an edge to SR nodes that require services that it can perform. Each service requesting agent (SR) is similarly connected by an edge to SP nodes that can provide the services that it requires. Each agent can communicate solely with agents that are connected to it by an edge.

The variables of the problem are held by the SPs, with assignments to the variables reflecting the actions (services) that they will perform. A solution to the problem will include an assignment to each of the variables. The solution's quality will be determined by the satisfaction of the SRs from the actions chosen by the SPs (the services assigned to be provided to them) and will be reflected in a global utility, which the agents aim to maximize. Thus, one set of agents (SPs) selects the actions that are performed, while the other set (SRs) evaluates the outcome of these actions.

Formally, a SOMAOP is a tuple $\langle SP, SR, S, PS, RS, X, D, U \rangle$, where $SP = \{SP_1, SP_2, \ldots, SP_n\}$ is a set of $n$ service providing agents and $SR = \{SR_1, SR_2, \ldots, SR_m\}$ is a set of $m$ service requesting agents.

The capabilities provided and requested as services are formalized as *skills*. The set of all skills is $S = \{S_1, S_2, ..., S_k\}$. Each $SP_i \in SP$ has a set of providable skills, $PS_i \subseteq S$. For each $s \in PS_i$, the SP has a workload $w_i^s$ that defines the amount of the skill it can provide as a service. For example, an ambulance can evacuate a limited number of casualties. For each skill $s \in PS_i$, the SP also has a work time function $t_i^s(w)$ that defines the time it takes to complete $w$ workload of this skill. The workload of a providable skill $s$ decreases when the SP schedules the skill to be provided as a service to an SR (providable skill $s$ is depleted when $w_i^s = 0$).

On the other hand, each $SR_j \in SR$ has a set of requested skills, $RS_j \subseteq S$. For each requested skill $s \in RS_j$, the SR has a workload $w_j^s$ that defines the amount of service required of the skill it requests. The workload of a requested skill $s$ decreases when an SP schedules to provide the skill as a service to the SR (requested skill $s$ is no longer required when $w_j^s = 0$). For each of its requested skills $s \in RS_j$ there is an optimal team size for performance capability, $q_j^{s^*}$, defining the number of SPs that are requested to cooperate simultaneously when performing the service (e.g., if a requested skill $s$ with $w_j^s = 2$ has $q_j^{s^*} = 2$, $SR_j$ will prefer two SPs to each schedule to provide half of the requested workload of $s$ simultaneously rather than a single SP to provide the full requested workload). Additionally, each requested skill $s$ has a maximal utility $u_j^{s^*}$, defining how much utility could be derived if the full service is completed immediately, with $q_j^{s^*}$ SPs sharing the workload of the service simultaneously. Lastly, each requested skill $s$ has a latest completion time $t_{max_j}^s$, after which the service is no longer required.

$X = \{X_1, X_2, ..., X_n\}$ includes sets of variables for each SP, i.e., for each service provider $SP_i$, $1 \leq i \leq n$, $X_i$ includes the set of variables $x_{i_1}, x_{i_2}, \ldots, x_{i_{\lambda_i}}$ representing the services that $SP_i$ will provide; $\lambda_i$ is the maximal number of services that it can perform. An assignment to $SP_i$'s variable $x_{i_a}$ is a service tuple $\langle SR_a, s_a, w_a, t_a \rangle$ representing the SR that the service will be provided to, the skill provided, the workload provided and the expected start time for performing the service, respectively. The order of the variables defines the order in which the agent will execute the services, i.e., $SP_i$ will first perform the service assigned to $x_{i_1}$, then the service assigned to $x_{i_2}$, etc.

$D = \{D_1, D_2, \ldots, D_n\}$ includes sets of variable domains such that $D_i$, $1 \leq i \leq n$, includes the set of domains $d_{i_1}, d_{i_2}, \ldots, d_{i_{\lambda_i}}$, which include the values that can be assigned to variables $x_{i_1}, x_{i_2}, \ldots, x_{i_{\lambda_i}}$ of $SP_i$ respectively (i.e., $d_{i_1}$ contains all of the service tuples that $SP_i$ can schedule to provide first). The domains can also include a non-service assignment in cases where a SP is purposefully not assigned to a service e.g., in cases when SPs need time to recharge.

A solution $\sigma$ to the SOMAOP is an assignment to each of the variables held by the set of SPs, of a value from its domain. The utility derived by a service requesting agent $SR_j$ from solution $\sigma$ is denoted by $U_j(\sigma)$. It is calculated as a function of the utility $SR_j$ will derive from the services scheduled for each requested skill $s \in RS_j$ as specified by $\sigma$, denoted $u_j^s(\sigma)$. $u_j^s(\sigma)$ is bounded by $u_j^{s*}$ and is affected by three factors: 1) The time the SR will spend awaiting service for $s$: the utility to be derived from the service will decrease with a latency penalty function, corresponding to the time the SR awaits service. 2) The amount of workload scheduled to be performed and its timing. 3) The performance capability of the SPs providing the service: the performance capability of $SR_j$'s requested skill $s$ is affected by the number of SPs that provide the workload of the service simultaneously [1]. This is denoted by the capability function, $Cap_j^s(q)$. The function can represent minimum required or maximum allowed numbers of agents by setting the capability to 0 for fewer agents, or by not increasing the capability when more than the maximum number of required agents share a service, respectively. $Cap_j^s(q)$ will reach its maxima at $q = q_j^{s*}$. We assume $Cap$ is weakly monotonically increasing in $q$.

$U(\sigma)$ defines the global utility derived from solution $\sigma$ and is a function of the utilities received by each of the SRs, i.e., $U(\sigma) = F(U_1(\sigma), U_2(\sigma), \ldots, U_m(\sigma))$. The goal of the agents in SOMAOP is to maximize the global utility function U.

## 3 ALGORITHMS FOR SOLVING SOMAOP

The general approach we take is to design iterative distributed algorithms in which the building blocks are existing methods for assigning service providers to services, used in the Operation Research literature, which we adapt to a distributed environment. Specifically, we will focus on two approaches: auctions [6, 18, 19, 26] and matching [4, 15, 16].

### 3.1 Repeated Parallel Auctions (RPA)

The RPA algorithm creates allocations of SPs to SRs by using a repeated auction process [18, 21, 23]. In each of the algorithm's iterations (a predefined number), an auction occurs between the SPs (sellers that offer providable skills) and the SRs (the buyers who offer bids on the skills they require). The auction begins with each SP sending a *service proposal* to its neighboring SRs for each of their joint skills (skills that the SP can provide and the SR requests). A service proposal from $SP_i$ to $SR_j$ for providable skill $s$ is composed of $SP_i$'s proposed workload for $s$ and the proposed service start time at which $SP_i$ proposes to begin providing $s$ to $SR_j$.

Upon receiving service proposals from its SP neighbors, each SR responds by sending *service requests* to the SPs that it would most want to provide it each of its requested skills. A service request from $SR_j$ to $SP_i$ for requested skill $s$ is composed of a $SR_j$'s requested workload for $s$, a requested start time for $SP_i$ to begin to provide $s$ to $SR_j$ and a bid value that expresses the utility it could derive from receiving $s$ with the workload requested at the start time requested.

Once all service requests for the iteration are sent, the SPs will attempt to create a schedule (each SP starts with an empty schedule in each iteration). The SP attempts to schedule the service requests in descending order of bid value. A schedule attempt for request $r$ succeeds if the completion time of the last scheduled request is earlier

---

**Algorithm 1** RPA: Service Provider $i$

1: **for** fixed number of iterations **do**
2:    Reset $w_i^s \ \forall \ s \in PS_i, X_i, t_{earliest} \leftarrow 0$
3:    $requests \leftarrow$ requests received from SRs in previous iteration, ordered by highest bid value
4:    **for** $r \in requests$ **do**
5:       $t_{earliest}^r \leftarrow$ earliest time after $t_{earliest}$ that $SP_i$ can begin serving $r$
6:       **if** $t_{earliest}^r \leq t_{start}(r)$ **and** $w_i^{s(r)} \geq w(r)$ **then**
7:          send proposal($SR(r), s(r), w_i^{s(r)}, t_{earliest}^r$)
8:          schedule($r$)
9:          Update $w_i^{s(r)}, t_{earliest}$
10:      **else**
11:         break
12:      **end if**
13:    **end for**
14:    **for** $SR_j \in$ neighbors **do**
15:      **for** $s \in PS_i \cap RS_j$ not proposed in current iteration **do**
16:          $t_{earliest}^{j,s} \leftarrow$ earliest time after $t_{earliest}$ that $SP_i$ can begin serving skill $s$ to $SR_j$
17:          send proposal($SR_j, s, w_i^s, t_{earliest}^{j,s}$)
18:      **end for**
19:    **end for**
20: **end for**

---

than the requested start time of $r$ and if the SP has enough workload left to provide it, given the services needed to fulfil the already-scheduled requests. The SP will continue to attempt to schedule requests until an attempt fails. The SP responds to a scheduled request with a service proposal to provide the service as requested. The SP responds to an unscheduled request with an updated service proposal including its updated remaining providable skills and workloads (the original skills and workloads, minus those needed for the scheduled requests) and its updated proposed service start time (the next time possible after the scheduled requests). This begins the next auction (iteration), and the process occurs again.

Algorithm 1 depicts the main procedure of the SPs. Initially, a SP will propose its neighboring SRs the earliest possible service start time as well as its entire workload per joint skill (lines 14-17, as there are no requests in the initial iteration). In later iterations, the SP creates a new schedule by responding to service requests received from SRs, ordered highest bid first (lines 3-4). The SP will schedule request $r$ for skill $s(r)$ if the earliest possible start time for the request, $t_{earliest}^r$, is earlier than (or equal to) the start time requested $t_{start}(r)$ and if the SP has enough workload $w_i^{s(r)}$ to fulfil the requested workload, $w(r)$ (line 6). If a request is scheduled, the SP proposes to provide the request (line 7). The SP then schedules the request in the next free slot in $X_i$, updates $w_i^s$ according to the requested workload and $t_{earliest}$ according to the expected completion time of the request (lines 8-9). If a request is not feasible, the SP stops the scheduling process (line 11). The SP responds to the unscheduled requests by sending the SRs new proposals to provide the skills after the scheduled requests, along with the workloads they will be able to provide at this later time (lines 14-17).

Algorithm 2 depicts the main procedure of the SRs. For each of its requested skills $s$, the SR will iterate over the proposals received from SPs that correspond with $s$, ordered by the quality of the

**Algorithm 2** RPA: Service Requester $j$

1: **for** fixed number of iterations **do**
2:     Reset $w_j^s \; \forall \; s \in RS_j$
3:     **for** $s \in RS_j$ **do**
4:         $proposals_s \leftarrow$ proposals received from SPs in previous iteration for $s$, ordered by quality of the proposed service
5:         **for** $p \in proposals_s$ **do**
6:             $w_s^a \leftarrow min\{w(p), w_j^s\}$
7:             $bid^{s,i} \leftarrow$ allocate$(SP_i(p), s, w_s^a, t_{earliest}(p))$
8:             send request$(SP(p), s, w_s^a, t_{earliest}(p), bid^{s,i})$
9:             Update $w_j^s$
10:            **if** $w_j^s = 0$ **then**
11:                 break
12:             **end if**
13:         **end for**
14:     **end for**
15: **end for**

proposals received (lines 3-5); i.e., maximal utility gain possible from the workload the SP has proposed, normalized by workload. For each proposal, the SR will allocate workload equal either to the workload proposed or the remaining unallocated workload requested (line 6). The bid value is then calculated by the SR and is sent as a request to the proposing SP. The remaining workload requested $w_j^s$ is updated according to the allocation (line 9). The allocation for a requested skill ends when the allocation has satisfied the request (lines 10-11) or there are no more proposals to allocate. The algorithm does not aim to allocate exactly $q^{s^*}$ SPs to $s$ but rather enough SPs to provide the workload requested.

*3.1.1 RPA convergence.* The convergence of RPA depends on how bids are calculated. There are several parameters that can be considered when a SR decides on a bid to be sent to a SP regarding a specific skill $s$, such as the expected satisfaction from the service, the expected starting time and the amount of workload potentially received for skill $s$ by other SPs prior to this starting time.

We will prove that when the following assumptions hold, the algorithm is guaranteed to converge. First, we assume that SRs will bid higher for an earlier starting time. Formally, for each SR, $SR_j$, for every two bids for the same skill $s$, $b_s$ and $b_s'$ with starting times $st_b$ and $st_{b'}$, $b_s > b_s'$ if and only if $st_b < st_{b'}$. We will further assume that a SP will schedule that skill $s$ is applied for serving some $SR$, at most once.

While our model is more abstract and there are other possibilities for calculating the bids, these assumptions hold in many realistic scenarios, where the quality of service provided by different agents is similar and the starting time is a critical parameter. One such scenario is the mass casualty incident problem we address in our experimental evaluation.

In our convergence proof we will use the following notations. We will denote by $TS$ the set of scheduled services that will not change in following iterations (the set can only grow as the algorithm proceeds). It includes scheduled services such as $ts_{ijs}^k$, which indicates that the $k$'th service provided by $SP_i$ will be to $SR_j$ on skill $s$, and that this fact will *not change later on*. When $TS$ includes all provided service requests, the algorithm converges. We will further denote by $hb^k$ the highest bid in iteration $k$ for a service that is not yet in $TS$ and by $hb_{SP}^k$, $hb_{SR}^k$ and $hb_s$ the SP and SR and skill that $hb$ corresponds to, respectively.

OBSERVATION 1. *In iteration $k + 1$, $hb_s^k$ of $hb_{SR}^k$ will be the first service that is not in $TS$ on $hb_{SP}^k$'s schedule.*

This is because SPs order services according to their bid sizes.

OBSERVATION 2. *The only way that $hb^{k+1}$ can be smaller than $hb^k$ is when the service that $hb^k$ corresponds to was added to $TS$.*

As Observation 1 notes, the highest bid would be the first one (apart from those in $TS$) handled by SPs. If $hb^k$ is not the highest bid in iteration $k + 1$, it can only be if it was added to $TS$ or if there is a larger bid sent in iteration $k + 1$.

LEMMA 3.1. *The number of consecutive iterations in which $TS$ does not grow is bounded by $2|SP| \cdot |S|$.*

PROOF. Under a given $TS$ set, the highest possible bid not yet in $TS$ will be added to $TS$ (since it is not surpassed, the $SP$ agent getting the bid will always give it a high priority, and the requesting agent gets it as soon as possible (otherwise, it would have given a higher bid). Thus, when discussing changes to $TS$ we can focus on looking at the highest possible bid that is not in $TS$ yet.

Initially $TS$ is empty. After the first iteration of the algorithm, $hb_{SP}^1$ schedules the corresponding request as a service. Since all $SRs$ in the first iteration considered the earliest possible arrival time of each $SP$, this bid will remain the highest and will not change. Thus, this scheduled service is added to $TS$.

In each of the following iterations, each SP has a schedule that was determined according to the bids it received in the previous iteration. According to Observation 1, following iteration $k$, in iteration $k + 1$, $hb_s^k$ will be scheduled first among all services not yet in $TS$ by $hb_{SP}^k$. Thus, either $hb^{k+1}$ is the same as $hb^k$, or, according to Observation 2, it was replaced by a higher bid. In both cases, $hb_{SP}^k$ will never submit an earlier arrival time to $hb_{SR}^k$ than the one it submitted at iteration $k + 1$ and therefore, it will never receive a bid for this service that is higher than the one it got for it in this iteration. Thus, the maximal number of different highest bids between consecutive additions to $TS$ is bounded by two iterations for each SP on each skill, i.e., $2|SP| \cdot |S|$. That is, after that number of iterations, it is guaranteed that one of those bids was the maximal possible one, and thus would be added to $TS$. □

PROPOSITION 3.2. *RPA converges within $2|SP|^2 \cdot |S|^2$ iterations.*

PROOF. According to our assumption, each SP serves an SR on a skill only once, i.e., the number of services that are added to $TS$ is bounded by $|SP| \cdot |S|$. From Lemma 3.1, the maximal number of iterations between each increment to the size of set $TS$ is bounded by $2|SP| \cdot |S|$. Thus, the maximal number of iterations before the algorithm converges is bounded by $2|SP|^2 \cdot |S|^2$. □

Our assumption that each $SP$ will serve a $SR$ agent with skill $s$ at most once can easily be relaxed to serving the $SR$ agent with some fixed number of times. The proof for Proposition 3.2 will only need to be slightly changed, multiplying our convergence bound by a fixed amount.

## 3.2 Distributed Simulated Repeated Matching Algorithm (DSRM)

The DSRM algorithm creates allocations of SPs to SRs by repeatedly simulating the outcome of a matching algorithm over time. Each agent (SP as well as SR) has an internal clock that begins at $t = 0$ and progresses throughout the DSRM algorithm. Each iteration considers a simulated time $t$ at which the agents execute an iterative Gale Shapley inspired many-to-one matching algorithm [15] to match SPs with SRs. The outcome of the matching algorithm is translated to service tuples by the SRs and scheduled by the SPs.

Once a SP is matched to a service to a SR, it can determine when it will finish providing the service, by calculating how long it will take to complete its assigned workload (using $t_i^s(w)$). This way it can also know what its remaining workload will be at a future time.

At each iteration, we simulate as though the previous allocations already happened, which means that the provided services and workload are updated as well as the internal clock of each agent (an explanation on how to distributedly synchronize the internal clocks to the next relevant start time in each iteration follows). In each iteration we want to make a decision for the next allocation at this time. This simulated matching process will end when there are no more SRs with remaining requested skills or no more SPs with remaining providable skills. The final schedule is the solution to the problem and will include the assignments that were "executed" during the simulated process in the order that they were simulated.

The iterative Gale Shapley inspired many-to-one matching algorithm is performed as follows. In each iteration, the SR calculates a bid value for each of its requested skills, for each neighboring SP that can provide the service. This bid value expresses the utility it could derive from receiving the service from the SP. The SRs share the bids with the SPs. Then, a distributed version of the Gale Shapley college admissions algorithm (DGS) [4] is executed to create a many-to-one matching. Each SR acts separately and simultaneously for each of its requested skills. Both the SPs and the SRs' requested skills rank one another according to the bid values. The SPs that have been matched will not take part in the next iteration. The SRs will take part in the next iteration if they have at least one requested skill $s$ that has not been matched with $q_j^{s^0}$ SPs (defined by $\min\{q_j^{s^*}$, number of SP neighbors with $s \in PS_i\})$, and there is at least one neighboring SP to provide the skill. The iterative matching algorithm ends when there are no SPs or SRs left to match. Note that the algorithm does not aim to allocate just enough SPs to provide the workload requested but rather $q^{s^*}$ SPs for each skill.

Algorithm 3 depicts the main procedure of the SPs. A SP initializes the times $t$, $t_{last}$ to 0 and its assignment for time $t$ as empty (line 1). The algorithm ends when the SP no longer has SRs to provide services to or no service left to provide (line 21). At each simulated time $t$, the SP updates its neighboring SRs regarding its providable skills by sending service proposals for each skill (lines 3-6) and receives bids from the SRs in response (line 9). These bids are used to rank the SRs in the DGS algorithm. Then, the DGS algorithm is performed iteratively until the SP has been matched or there are no SRs left to match with (lines 11-13). Thereafter, the SP will receive an allocation from its matched SR (if one exists) (line 15). Lastly, the simulation time is updated, the SP assigns the completed portion of

---

**Algorithm 3** DSRM: Service Provider $SP_i$

1: $t_{last} \leftarrow 0, t \leftarrow 0, allocation\_t \leftarrow$ null
2: **repeat**
3:     **for** $SR_j \in$ neighbors **do**
4:       **for** $s \in PS_i \cap RS_j$ **do**
5:         $t_{earliest}^j \leftarrow$ earliest time after $t$ that $SP_i$ can begin serving $SR_j$
6:         send proposal($SR_j, s, w_i^s, t_{earliest}^j$)
7:       **end for**
8:     **end for**
9:     Receive bids from SRs and rank them accordingly
10:    Update neighbors
11:    **while** |neighbors| > 0 **and** has no match **do**
12:       Run *Distributed Gale Shapley*
13:       neighbors $\leftarrow$ neighboring SRs that require some skill from $PS_i$ and haven't completed their matching
14:    **end while**
15:    $assignment\_t \leftarrow$ receive assignment from matched $SR$
16:    $t_{last} \leftarrow t$
17:    $t = minimal\_SP\_finish\_time()$
18:    $pa \leftarrow$ partial $assignment\_t$ complete in $t - t_{last}$
19:    schedule($pa$)
20:    Update $w_i^{s(assignment\_t)}$ according to $pa$
21: **until** |neighbors| = 0 **or** $PS_i = \emptyset$

---

**Algorithm 4** DSRM: Service Requester $SR_j$

1: $t_{last} \leftarrow 0, t \leftarrow 0, assignment\_t \leftarrow \emptyset$
2: **repeat**
3:     Receive service proposals from SPs
4:     Update neighbors
5:     Calculate bids for all SP neighbors and rank them accordingly
6:     **for** $SP_i \in$ neighbors **do**
7:       **for** $s \in RS_j \cap PS_i$ **do**
8:         send request($SP_i, s, w_j^s, t_{earliest}^i, bid_{s,i}$)
9:       **end for**
10:    **end for**
11:    **while** |neighbors| > 0 **and** has $s \in RS_j$ that has not been matched with $q_j^{s^0}$ SPs **do**
12:       **for** $s \in RS_j$ **do**
13:         $q_j^{s^0} \leftarrow min\{q_j^{s^*}, |$SP neighbors with $s$'s skill$|\}$
14:       **end for**
15:       Run *Distributed Gale Shapley*
16:       neighbors $\leftarrow$ all neighboring SPs that can provide some $s \in RS_j$ and haven't been matched
17:    **end while**
18:    $assignment\_t \leftarrow assign\_providers$(matched SPs)
19:    $t_{last} \leftarrow t$
20:    $t = minimal\_SP\_finish\_time()$
21:    **for** $SP_i \in assignment\_t$ **do**
22:       $pa \leftarrow$ partial assignment complete in $t - t_{last}$ by $SP_i$
23:       Update $w^s(assignment\_t)_j$ according to $pa$
24:    **end for**
25: **until** |neighbors| = 0 **or** $RS_j = \emptyset$

---

the service to its schedule according to the elapsed time ($t - t_{last}$), and its remaining providable skills are updated (lines 17-20).

Algorithm 4 depicts the main procedure of the SRs. At first, similarly to the SPs, a SR initializes the times $t$, $t_{last}$ to 0 and its allocation for time $t$ as empty (line 1). The algorithm will end when the SR has no more requested skills or when the SR no longer has SPs that can provide its requested skills (line 25). At each time $t$ in which the simulation is performed, the SR receives the SP's service proposals (line 3), calculates bids (as described in the following sub-section) for each of its neighbors per skill they have that the SR requires and sends service requests to the SPs (lines 5-8). The calculated bids are used to rank the SPs in the DGS algorithm. Then, the DGS algorithm is performed iteratively for each of the requested skills simultaneously, until each skill $s \in RS_j$ has been matched with $q_j^{s^0}$ SPs (defined by $\min\{q_j^{s^*}$, number of SP neighbors with $s \in PS_i\}$), or there are no SPs left to match with (lines 11-16). The SR allocates services to be performed by the SPs by dispersing the load evenly between the matched SPs, considering their available providable skills (line 18). Lastly, the simulation time is updated and the SR's remaining requested skills are updated according to the workload that has been completed in the elapsed time ($t - t_{last}$) (lines 20-23).

To find the minimal next simulation time (line 17 in algorithm 3, line 20 in algorithm 4), we use a simple distributed algorithm (inspired by [10]). Each agent (whether a SP or SR) holds a minimal time (for a SP it will be initialized as the completion time of its allocation; for a SR it will be initialized as the earliest completion time of its allocated SPs) and sends this time to its neighbors. Each agent receives its neighbors' messages and saves the minimal time. When the minimal time of an agent is revised, it is sent to its neighbors. This algorithm (that finds the next minimal simulation time) will converge in $O(d)$ iterations (d being the diameter of the communication graph), as agent $a$ that has the true minimal time will surely never change it. Therefore, at most, the message will have to reach the furthest agent from $a$ in the graph.

*3.2.1 DSRM Properties.* In order to establish the following property we first assume that there is a minimal amount of workload that an SP will perform when assigned to apply some skill, serving some SR. We note this minimal fraction of workload by $\epsilon$.

PROPOSITION 3.3. *DSRM converges to a solution in a pseudo-polynomial number of iterations.*

PROOF. According to our assumption, the number of possible assignments to apply a skill for some SR is bounded by the number of SRs (m) times the number of skills (k) times the maximal number of fractions of workload ($\frac{w}{\epsilon}$), where $w$ is the maximal workload requested for any skill. Since in every iteration of the algorithm at least one SP is assigned to perform some skill in order to serve some SR, and this assignment is not changed in later iterations, the number of iterations is bounded by: $n \cdot m \cdot k \cdot \frac{w}{\epsilon}$. Thus, the number of iterations before the algorithm converges is pseudo polynomial. □

PROPOSITION 3.4. *The quality of the solutions found by DSRM as a function of the number of iterations is monotonically increasing.*

PROOF. The solution is incrementally built. After beginning empty, at each iteration, at least one assignment of a SP to perform a skill for an SR is added to the partial solution. Each such assignment has positive utility. Therefore the quality of the solution (which is the sum of the utilities derived for each such assignment) is increasing with each iteration. □

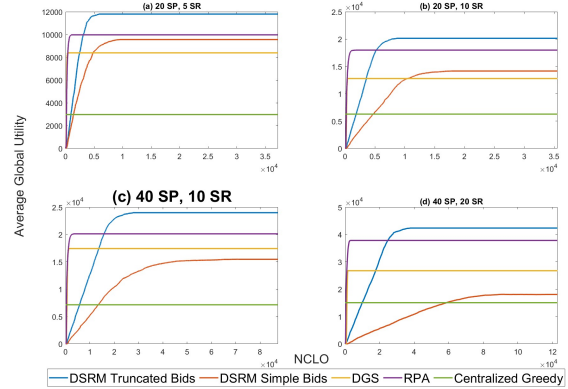

**Figure 1: Abstract Simulator**

*3.2.2 Calculating DSRM Algorithm Bids.* We propose two functions for calculating bidding values:

**Simple** assigns each SP neighbor a bid value for each of its $s \in RS_j$ that represents the utility the SR would derive if the SP was to provide as much of $s$ as possible to the SR disregarding other SPs' abilities and the Cap function.

**Truncated** assigns positive values only to a number of SPs for each of its $s \in RS_j$. The number of SPs that will receive positive bids is equal to $q_j^{s^0}$. The SR chooses the $q_j^{s^0}$ SPs with the earliest expected start times and assigns to each of them a value that represents the marginal utility it should receive, taking into account the SPs that could arrive before it as well as the effect of the Cap function (more details can be found in the supplementary material).

## 4 EXPERIMENTAL EVALUATION

To evaluate the proposed algorithms' performance, we created two different simulators. The first simulates the coordination between SPs and SRs of an abstract SOMAOP, with an objective of maximizing the global utility function. The second simulates a specific and realistic instance of SOMAOP, the coordination between medical units (SPs) and disaster sites (SRs) in a Mass Casualty Incident (MCI) setting with an objective of minimizing the number of casualties with a low survival probability [25].

All results presented are averages of solving attempts of the 50 simulated problems, by the algorithms. Figure 1 and 2 present the global utility as a function of Non-Concurrent Logic Operations (NCLOs) [20, 24, 34] for four scenarios in the abstract simulator and the MCI simulator, respectively. Each scenario has a different *magnitude*, or ratio of SP size to SR size. The scenarios in our experiments had 40 and 20 SPs and a magnitude of 4 : 1 and 2 : 1. In each scenario we compared five algorithms: RPA, DSRM using the simple bid function, DSRM using the truncated bid function, the Distributed Gale Shapley College Admissions algorithm (DGS) as a one-shot schedule and a centralized greedy algorithm. In the centralized greedy algorithm pairs of SP and a requested skill of an SR are selected and scheduled sequentially, ordered according to the maximal utility per workload. The algorithm continues until there are no more SRs with remaining requested skills that can be served by the SPs.

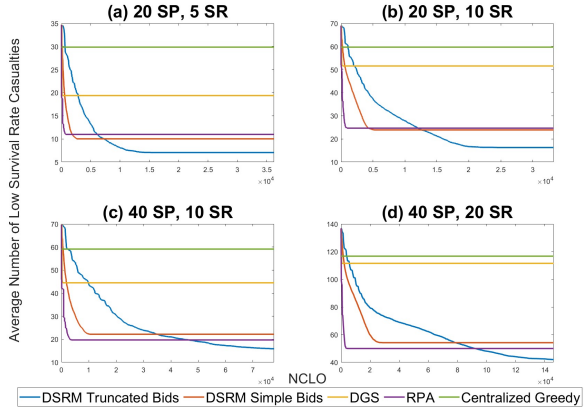

Figure 2: MCI Simulator

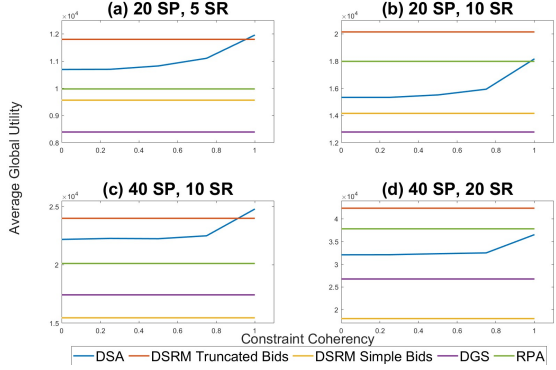

Figure 3: Constraint coherence Abstract Simulator

### 4.1 Comparing SOMAOP Algorithms

The results presented in Figure 1 show a clear and consistent advantage of the version of DSRM that uses the truncated bid function. In comparison to DSRM, RPA converges earlier, but to solutions with a lower global utility. The GS algorithm converges fastest, since it only performs a single shot schedule. The DSRM version that uses a simple bid function produced solutions with a lower utility on average than the results produced by the version that used truncated bid. Moreover, its runtime was longer due to the larger number of iterations it performs in each execution of the Distributed Gale Shapley algorithm. As the amount of SPs increases, the runtime of DSRM increases (regardless of the type of utility being used). In contrast, in RPA the convergence time is faster than DSRM regardless of the amount of SPs. The solutions that DGS produces have lower utility than the utility of solutions produced by DSRM with a simple bid function when there are 20 SPs and higher when there are 40. It seems that this is the effect of DSRM's readjustment each time SPs are planning to end a service. When there are many SPs, such adjustments occur often. This results in SPs abandoning their services for higher bidders, meaning their time is wasted and thus the utility decreases. In these cases, DGS performs better despite creating a single-SR-schedule for the SPs, as the scheduled services are completed at the earliest available time with no delay. DSRM with a truncated bid is not fazed by the amount of SPs as the bids are calculated in a way that is less sensitive to changes. The centralized greedy algorithm is shown as a horizontal line describing the average final utility of the algorithm (as opposed to utility over NCLOs). This approach produced lower utility results in all of the problem sizes shown.

Figure 2 presents similar results of the algorithms solving MCI problems. Again, DSRM using truncated bid yields the highest quality results, and RPA converges fast regardless to the amount of SPs. However, DSRM with simple bid converges much faster on this simulator. The reason is that there are strict ordering constraints between skills applied by SPs in this simulator, i.e., medical treatment must be given before evacuation to the hospital. Thus, optional outcomes are ruled out and the size of the solution space is much smaller than that of the abstract simulator. This is also the reason for the clear difference between the results of DSRM with a simple bid function and the results of DGS.

### 4.2 Comparing SOMAOP and DCOP Algorithms

To compare SOMAOP algorithms with DCOP algorithms, we need to describe how an instance of SOMAOP is modeled as a DCOP (similar to how multi agent task allocation problems were modeled as DCOPs in [2]). First, we note that in a DCOP there is only one type of agents, i.e., the DCOP agents are the SPs and there are no agents representing the SRs. Thus, in DCOP, each of the SP agents must be able to communicate with the other SP agents. A SP agent neighbors another SP agent if they can both provide the same skill to a SR. Additionally, besides holding variables and variable domains as they do in SOMAOP, the SPs must also hold the constraint information (the utility derived from different combinations of decisions regarding service providing). Moreover, many DCOP algorithms require the SP to correctly calculate the utility from an assignment to its variables, thus, it must also know the assignments of its neighboring SPs.

*Information coherence* of a DCOP as the extent to which each agent is aware of the characteristics of the DCOP (i.e., other agents' assignments or the constraints of the problem). High coherence is associated with the agents having a more complete and intelligible awareness of the state of other agents in the DCOP and the constraints among them. Low coherence is associated with the agents having an incomplete and unintelligible awareness of the DCOP elements. One possible reason for low coherence is the attempt to preserve agents' privacy. Low coherence may also be associated with imperfect communication [27, 36]. We distinguish two types of coherence, inspired by [11]'s definitions of privacy guarantees in DCOPs: **1)** Assignment coherence: The extent to which an agent is aware of the assignments chosen by other agents to their variables. **2)** Constraint coherence: The extent to which an agent is aware of the cost incurred by the constraints in the problem.

The separation of the responsibilities between the two sets of agents in SOMAOP such that only the SRs need to be able to evaluate possible solutions and be aware only of the utility calculation regarding their own set of requested skills, allows the SPs to focus only on their own current state. All the SP needs to know is the information regarding the utility derived from its own choice of assignments. This information is delivered to the SP by its neighboring SRs in SOMAOP. Thus, the required information coherence in SOMAOP is negligible for both forms of coherence defined.

In DCOP algorithms, in order for the agents to be able to evaluate the quality of their value assignments, they must know all the constraints they are involved in. Thus, the required constraint coherence of standard DCOP algorithms is high. In terms of assignment coherence, the SOMAOP model eliminates the need for the SPs to know of other SPs' assignments as SRs are the only ones that must see the "bigger picture" of assignments in the system. Therefore, the SOMAOP algorithms do not require assignment coherence for the SPs. DCOP algorithms, on the other hand, requires the agents to know all neighborequrents assignments, i.e., the assignment coherence requirement in DCOP is also high.

The high requirement for the coherence of the information held by the SP agents in DCOP violates the essential distributed properties which are preserved in the SOMAOP. If each SP has access to all constraints regarding each of the neighboring SRs as well as access to all of the other SPs' assignments, perhaps a centralized approach is equivalently appropriate.

Using the same scenarios as in the experiments presented above, we compare the SOMAOP algorithms – RPA and DSRM – to DCOP's DSA [33] and Max-Sum [12, 35]. We begin with the DSA: we used DSA-C with a probability $p = 0.7$ for replacing a value assignment. In each iteration each agent selected a random variable $x_i$ to which it considered whether to replace its assignment to the best alternative. To evaluate the relation between information coherence and the quality of solutions reported by DSA, we limited the information coherence of the agents performing the algorithm and compared the results to the outcomes of the SOMAOP algorithms.

To limit information coherence we define $p_c, p_a \in [0, 1]$, which determine the amount of information an agent knows regarding its neighbors' constraints and assignments respectively. For example, $p_c = 0.5$ translates to a 50% chance of an agent being aware of the cost incurred by a specific constraint in the problem.

Figures 3 and 4 present the results for constraint coherence ($p_a = 1$), and assignment coherence ($p_c = 1$), respectively. The results in Figure 3 show that for problems with a 4:1 ratio between SPs and SRs respectively, DSA outperforms DSRM with a truncated bid function when the Constraint Coherence is above $p_c = 0.75$. In problems with a 2:1 ratio, even with $p_c = 1$, our algorithms provide a better average final global utility. The results presented in Figure 4 show similar outcomes. For problems with a 4:1 ratio between SPs and SRs respectively, DSA outperforms DSRM with a truncated bid function only when the Assignment Coherence is above $p_a = 0.75$ when there are 20 SPs, and above $p_a = 0.5$ for 40 SPs. In problems with a 2:1 ratio, even with $p_a = 1$, our algorithms provide a better average final global utility. Similar results were also shown in the MCI simulator. The results show that although the DCOP framework can be used to solve SOMAOPs, it requires a high information coherence from the agents in order to achieve similar (or worse) results than those of SOMAOP algorithms.

The Max-Sum algorithm operates on a bipartite factor graph [12, 35]. This characteristic makes Max-Sum seem like a natural choice for solving service-oriented multi-agent problems. However, when used to solve problems whose inherent structure is of a bipartite graph (including service providers and service requesting agents), the algorithm fails to overcome its inherent symmetry and performs poorly [9, 31]. Additionally, since the constraints held by SRs in SOMAOP can involve a large number of SPs, i.e., they are constraints with high arity, the function-nodes in Max-Sum must use exponential runtime in order to generate messages.

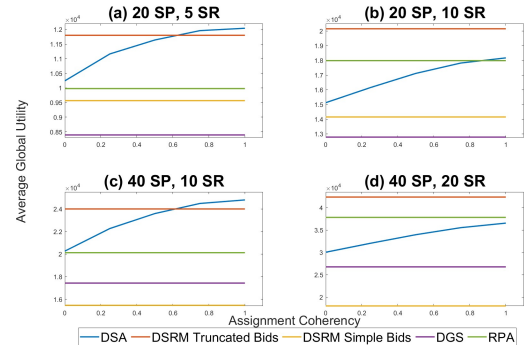

Figure 4: Assignment Coherence Abstract Simulator

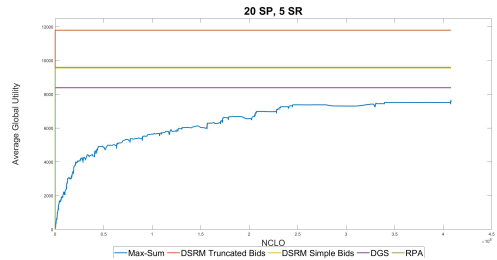

Figure 5: Max-Sum Comparison in the Abstract Simulator

To see how well Max-Sum can handle instances of SOMAOP, the algorithm was implemented on the abstract simulator and compared with our proposed algorithms. Here too, we implemented an iterative approach (which significantly outperformed a single shot approach) in which the solution was built incrementally by performing Max-Sum in each iteration in order to allow the SPs to select their next action.

Figure 5 presents the average quality of the solutions produced by the algorithms, solving 30 problems, with 20 SPs and 5 SRs. The exponential runtime of Max-sum prevented us from experimenting with larger problems. The results indicate that Max-sum produces solutions with far lower quality than the SOMAOP algorithms.

## 5 CONCLUSIONS

Many realistic distributed problems include service requesters and service providers. In the last two decades, distributed optimization problems have been represented and solved using the DCOP model and algorithms, which are not suitable for representing the two types of agents in service oriented multi agent optimization problems. Additionally, they require high information coherence, which is often unwanted or simply unrealistic in the environments of real-life problems. We proposed SOMAOP, a novel model for representing such problems and algorithms for solving them. The algorithms use well studied allocation methods as building blocks, and update the agents' estimations (bids) of utility from the services available following each iteration. Our empirical results demonstrate the advantages of the proposed iterative processes for solving this type of problem.

## ACKNOWLEDGMENTS

This work was supported in part by Israel Science Fund (ISF) grants 1965/20 and 3152/20, as well as funding from the Bi-national Science Foundation (BSF).

## A EXPERIMENTAL SETUP

In the abstract simulator, the set of skills consisted of 4 skill types. For each skill type, a work time was drawn from a half-normal distribution with $\mu = 1$ and $\sigma = 1$, and was constant for all SPs that could provide the skill as a service. For both the SPs and SRs, we randomly chose how many and which of the skills they would be able to provide or require, respectively. For each skill that was chosen, a workload was randomly selected. The work time, as a function of the workload, was linear and defined by $t_i^s(w) = t_i^s \cdot w$, where $t_i^s$ was a constant per skill and per SP. In addition, both SPs and SRs had a location on a grid. The SPs had the ability to travel in the grid, with the distance from them to their destination delaying the performance of a service to an SR. An SR's location remained constant. Each SR had a latest completion time, which was constant for all of its requested skills. The maximal utility of each of a SR's requested skills, $u_j^{s^*}$ was drawn from $U(750, 2000)$. For each scheduled execution of a skill, the requester derived partial utility. The utility of a requested skill deteriorated linearly over the time when no SPs were scheduled to provide it (so later execution resulted in lower utility). The optimal requested team size, $q_j^{s^*}$ for each of a SR's requested skills was randomly chosen to be either 1 or 2. The Cap function affected the quality of the execution according to the number of SPs providing a service simultaneously: $Cap_j^s(q) = min\{\frac{q}{q_j^{s^*}}, 1\}$.

In the MCI simulator[1], each disaster site (SR) had a number of casualties, randomly selected between 5 and 10. The casualties were classified to 3 levels of injury severity (urgent, medium, and non-urgent). Each of the casualties needed to receive on-field medical treatment, to be uploaded to an ambulance and transported to a hospital. Since there are 3 levels of severity and 3 services each, there are 9 skill types. There were 3 types of service providing medical units: Advanced Life Support (ALS) ambulances, Basic Life Support (BLS) ambulances, and motorcycles. Each medical unit (SP) can provide a set of skills according to its type. In addition, according to its type, each medical unit has a capacity of victims it could evacuate from the scene (e.g., ability to evacuate only 2 casualties) [3]. The medical units and the disaster sites had a location on an 80 km x 80 km grid. The medical units had the ability to travel at $60\frac{km}{h}$. Each casualty had a survival function that deteriorated over time as a function of its injury severity [7]. Thus, delays in the execution of services decreased the probability that the casualties will survive. The objective was to minimize of the total number of casualties with a low probability of survival (under 40%).

The MCI in multi disaster sites simulator consists of disaster sites, casualties, medical units, and hospitals[2].

Disaster sites differ in their location, the number of casualties and the severity of their injuries. To determine the number of casualties and their severity, we divided the disaster sites into three levels of importance [25]. According to the disaster site importance we determines the casualties amount and their severity of injuries. The different probabilities for each listed in table 1.

---

[1]we designed a simulator under the supervision of an expert in emergency medicine; detailed information can be found in the supplementary material in Appendix A.
[2]The simulator where designed under the supervision of an expert in emergency medicine

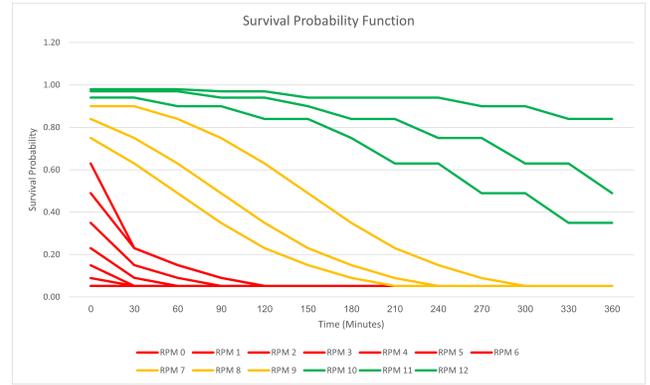

Figure 6: Survival probability function (Sacco et. al 2005)

Each casualty had a triage classification (urgent, medium and non urgent), and an RPM which was determined randomly uniformly according to the casualty's triage. The various RPM are detailed in 6 when red is for urgent triage, yellow for medium triage, and green for non urgent triage. Each of the casualties needed to receive on-field medical treatment, to be uploaded to an ambulance and transported to a hospital. The time needed for medical treatment and uploading for each RPM lengthens over time [7] as documented in 7. The time required for evacuation is based on the time it takes to travel from the disaster site location to the nearest hospital.

There are 3 types of medical units: Advanced Life Support (ALS) ambulances, Basic Life Support (BLS) ambulances, and motorcycles. Each type has different capacities and capabilities. An ALS can perform all activities: medical treatment, uploading and evacuation for all types of casualties (urgent, medium and non-urgent). A BLS can perform all activities, but only for medium and non urgent casualties. Motorcycles can provide only medical treatments for all types of injuries. The medical unit's capacities depend on the triage of the evacuated casualties. Urgent casualties are evacuated lying down, and medium and non-urgent casualties are evacuated while sitting. As a result, there are different capacities for each medical unit. The various capacities are listed in 2. The proportions between the types of medical units are based on the proportion between the numbers of medical units operating in the city of Beer Sheva. The location of the disaster and the congestion on the roads have a strong impact on the travel time. Therefore, we used the average travel time, which is 60 km/h for each medical unit. For each medical unit and for each site, we selected a random location within an 80 km x 80 km grid, assuming one hospital with infinite capacity.

## B COMPARING SOMAOP AND DCOP INFORMATION COHERENCE

### B.1 Constraint Coherence

To limit constraint coherence of the information held by agents in a DCOP, we define that with some probability, $p_c$, a SP will know the utility resulting from a combination of assignments, i.e., an entry in the utility table, and will substitute it with an unwanted outcome. The unwanted outcome was chosen to be 0 in the abstract simulator and the maximal number of low survival rate casualties for the MCI simulator. This means that the agents preferred to stray away from decisions that they were uncertain about. When $p_c = 1$, the SP has full constraint coherence and therefore knows the utility of each constraint; when $p_c = 0$, the SP has no constraint coherence and therefore does not know the utility of any of the constraints. In the following figures, the assignment coherence was kept constant with $p_a = 1$.

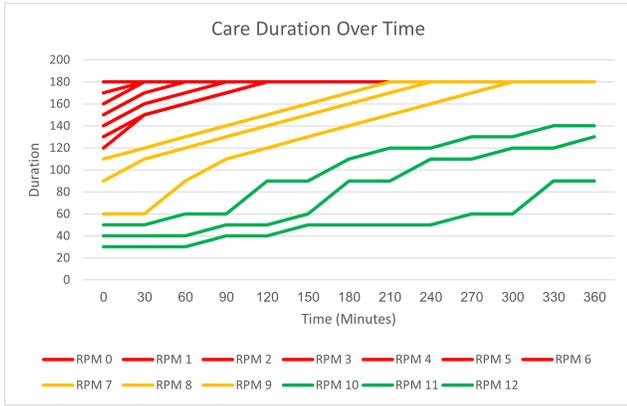

Figure 7: Care Duration Over Time

Table 1: Disasters sites probabilities

| Disaster Site Type | Casualties amount | Triage | Triage Probabilities | Type Probability |
|---|---|---|---|---|
| 1 | 5 | Urgent | 0.6 | 0.2 |
|   |   | Medium | 0.2 |   |
|   |   | Non Urgent | 0.2 |   |
| 2 | 8 | Urgent | 0.2 | 0.2 |
|   |   | Medium | 0.4 |   |
|   |   | Non Urgent | 0.4 |   |
| 3 | 10 | Urgent | 0 | 0.6 |
|   |   | Medium | 0.3 |   |
|   |   | Non Urgent | 0.7 |   |

Table 2: Medical unit type probabilities

| Unit Type | Probability | Activities | | | Casualties' capacity | | |
|---|---|---|---|---|---|---|---|
|   |   | Treatment | uploading | transfer | Urgent | Medium | NU[3] |
| Als | 0.2 | V | V | V | 2 | 0 | 0 |
|   |   |   |   |   | 1 | 2 | 0 |
|   |   |   |   |   | 1 | 1 | 1 |
|   |   |   |   |   | 0 | 0 | 6 |
|   |   |   |   |   | 0 | 4 | 0 |
|   |   |   |   |   | 1 | 0 | 3 |
| Bls | 0.6 | V | V | V | 0 | 2 | 0 |
|   |   |   |   |   | 0 | 1 | 1 |
|   |   |   |   |   | 0 | 0 | 3 |
| motorcycle | 0.2 | V | X | X | 0 | 0 | 2 |
|   |   |   |   |   | 0 | 2 | 0 |
|   |   |   |   |   | 0 | 1 | 1 |
|   |   |   |   |   | 1 | 0 | 0 |

### B.2 Assignment Coherence

To limit assignment coherence of DCOP agents, we define that with some probability, $p_a$, a SP will not know the assignments of a neighboring SP. When a SP does not know the assignment of a neighbor, he acts as though the neighbor does not exist by considering only constraints that the neighbor is not involved in.

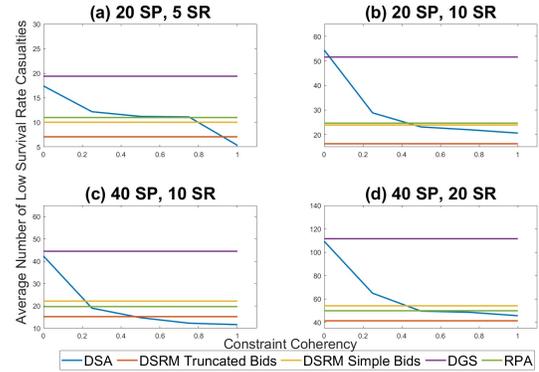

Figure 8: Constraint Coherence MCI Simulator

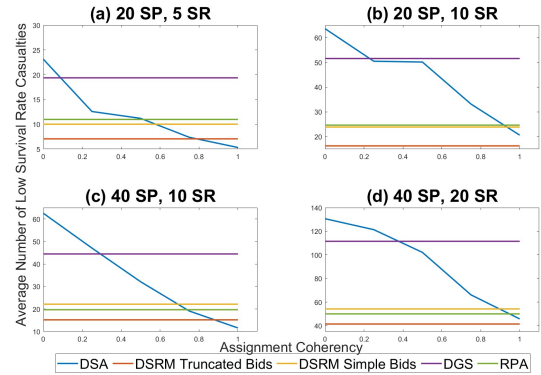

Figure 9: Assignment Coherence MCI Simulator

When $p_a = 1$, the SP has full assignment coherence and therefore knows the assignments of all its neighboring SPs; when $p_a = 0$, the SP has no assignment coherence and therefore does not know of any of its neighbors assignments. In the following figures, the constraint coherence was kept constant with $p_{so} = 1$.

### B.3 MCI Information Coherence Outcomes

The results were shown in 8 is similar to the tendency shown in the abstract simulator. The only difference is that when the ratio is 4:1 between SPs and SRs, DCOP achieved a better outcome only when the Constraint Coherence was above $p_c = 0.75$ for 20 SPs and above $p_c = 0.5$ for 40 SPs. In problems with a 2:1 ratio, even with $p_a = 1$, our algorithms provide a better solutions.

The results presented in Figure 9 show a similar outcome as with the constraint coherence. For problems with a 4:1 ratio between SPs and SRs respectively, the DCOP will only achieve a better outcome than that of DSRM with a Truncated Bid function when the Assignment Coherency is above $p_a = 0.75$. In problems with a 2:1 ratio, even with $p_a = 1$, our algorithms provide a better average final global utility.